\documentclass[a4paper,aps,prd,onecolumn,preprintnumbers,showpacs,nofootinbib]{revtex4}
\usepackage{amsmath,amssymb,graphics,epsfig,subfigure}
\usepackage{color}

\def\blue{\color{blue}}
\def\red{\color{red}}

\begin{document}
\thispagestyle{empty}

\begin{center}

\title{Establishing a universal relation between gravitational waves and black hole lensing}

\date{\today}
\author{Shao-Wen Wei\footnote{E-mail: weishw@lzu.cn},
        Yu-Xiao Liu\footnote{Corresponding author. E-mail:liuyx@lzu.edu.cn}}


 \affiliation{Institute of Theoretical Physics, Lanzhou University,
           Lanzhou 730000, China}

\begin{abstract}
Black hole lensing and gravitational waves are, respectively, closely dependent of the property of the lens and radiation source. In this letter, a universal relation between them is established for a rotating black hole acting simultaneously as a lens and a gravitational wave source, in an asymptotically flat spacetime. The relation only relies on the lens geometry and observable, while is independent of the specific nature of the black hole. Therefore, the possible gravitational wave sources could be located with modern astronomical instrument from the side of the lensing without knowing the specific nature of the black hole lens. Moreover, the low bound of the frequency of the gravitational waves can also be well determined.
\end{abstract}


\maketitle
\end{center}

\section{Introduction}
The theoretical prediction of the black holes in general relativity arouses a great interest in studying strong gravitational physics. Black hole is also generally believed to be a good test bed for both gravity and quantum mechanics. Through the indirect astrophysical observation, it implies that there exist many candidates of black holes in our universe, especially in the center of galaxies.

In general, each black hole is endowed with a Hawking temperature \cite{Hawking1,Hawking2}, determined by its unique nature. However, for a black hole with the same mass of the sun, its Hawking temperature is of $10^{-8}$ K, which is very difficult to directly observe using the thermodynamic method. Fortunately, a black hole's existence can sometimes be inferred by observing its gravitational interactions with its surroundings. If the black hole background is stable and ignoring the impact of the small mass particle on the background, the gravitational effects can be well understood with the geodesics of the test particle. Therefore, the geodesic plays a significant role in studying the gravitational effects.

Especially, the unstable circular null geodesic is closely related to the astronomical observation. The corresponding physics includes the quasinormal mode (QNM) frequencies, lensing in strong gravitational fields, and so on. The QNM frequencies \cite{Kokkotas,Nollert,Berti} were considered to be the resonances of a classical scattering problem, for example the scattering of particles in the vicinity of a black hole with the boundary conditions that only purely outgoing waves at infinity and ingoing waves at the event horizon. Analytically solving the QNM frequencies through the scattering equation is not an easy work. However, in the eikonal limit, Mashhoon \cite{Mashhoon} presented an analytical method to obtain them. Following this method, the QNM frequencies were thought to be massless particles propagating along the unstable circular orbit and slowly leaking out to infinity \cite{Press,Goebel,Ferrari,Berti2}. This was thoroughly studied by Cardoso \emph{et. al.} in Ref. \cite{Cardoso}. The real part of the QNMs is related to the angular velocity of the unstable circular null orbit, and the imaginary part was related to the Lyapunov exponent that determines the instability time scale of the orbit:
\begin{eqnarray}
 \omega_{\text{QNM}}=\Omega_{c}m-i(n+1/2)|\lambda|.\label{One}
\end{eqnarray}
This result was proved to be valid for not only the static, spherical spacetime, but also the equatorial orbits in the geometry of rotating black holes solution. A few years ago, the gravitational waves (GWs) were closely linked to the QNM frequencies. In particular, lowest lying modes of QNM would give a better understanding of the properties of the gravitational wave signal. Thus, one could keep in mind that the QNMs dominate the gravitational wave radiating from a black hole. And the study of QNMs will put forward to directly test of the GWs. However, GWs detection is still a grand challenge due to the lack of the strong GWs source with proper frequency.

On the other hand, black hole lensing is more easily observed by modern astronomical instrument, such like the very long baseline interferometry with a reachable resolution of $\sim 10\; \mu arcsec$. In the strong gravitational fields, the black hole lensing is related to the unstable circular orbit (for a recent review see \cite{Bozza}). This, to some extent, implies that there is an elegant relation between the black hole lensing and GWs. This conjecture was first made by Decanini and Folacci \cite{Decanini}. And subsequently, the relation was found by Stefanov, Yazadjiev, and Gyulchev \cite{Stefanov} for a static, spherically symmetric black hole in an asymptotically flat spacetime. Considering that a black hole acts simultaneously as a gravitational lens and a source of GWs, then through examining the property of the black hole lensing, one could have the opportunity to obtain the characteristic frequencies of emission of GWs of the observed objects. This will help us solving the difficulty that how to determine the localization of the GW sources \cite{Stefanov}.

However, non-rotating objects are more an exception than a rule in our universe, in particular, observations show that rotating black holes most likely exist \cite{Narayan}. Much work shows that the dimensionless spin may adopt a larger value, or even approach its practical limitation $a/M=0.998$ \cite{Thorne,Narayan2}. Thus, in order to determine the nature of the black holes or explore the GWs using such relation \cite{Stefanov}, one faces the problem that the background should be generalized to a stationary, axis-symmetric spacetime. And this is the aim of the current letter. We also apply the relation to the Kerr-Newman (KN) black hole as an example, which shows an enormous help for us to determine the nature of a black hole and the detection of GWs if the rotating black hole acts simultaneously as a gravitational lens and a source of GWs. The low bound of the frequency of the GWs can also be well determined.

\section{Black hole lensing and quasinormal mode frequency}
Generally, the petrov-type D metric can be written as
\begin{eqnarray}
 ds^{2}=&-&\frac{\Delta}{\rho^{2}}\bigg(dt+P(\theta)d\phi\bigg)^{2}
      +\frac{\rho^{2}}{\Delta}dr^{2}+\rho^{2}d\theta^{2}\nonumber\\
      &+&\frac{\sin^{2}\theta}{\rho^{2}}\bigg(adt+R(r)d\phi\bigg)^{2},\label{Kerr}
\end{eqnarray}
where $P(\theta)$ and $R(r)$ are only the functions of $\theta$ and $r$, respectively. Imposing the boundary condition, this metric can describe an asymptotically flat black hole with or without the spin parameter. In such spacetime, there are two Killing fields $\xi_{t, \phi}=\partial_{t, \phi}$ related to two conserved constants along the geodesics,
\begin{eqnarray}
 E&=&-g_{\mu\nu}\xi_{t}^{\mu}p^{\nu},\\
 L&=&g_{\mu\nu}\xi_{\phi}^{\mu}p^{\nu},
\end{eqnarray}
with $p^{\mu}$ the four-momentum of a test particle. $E$ and $L$ represent the energy and angular momentum of the particle. In this letter, we only focus on the equatorial plane ($\theta=\pi/2$), the reduced metric can be expressed as
\begin{eqnarray}
 ds^{2}=-A(r)dt^{2}+B(r)dr^{2}+C(r)d\phi^{2}-D(r)dtd\phi.\label{reducemetric}
\end{eqnarray}
Comparing with (\ref{Kerr}), the metric functions can be obtained. Then the motion of a photon could be derived following Chandrasekhar \cite{Chandrasekhar} with all the coordinates and parameters adimensionalized by the black hole mass $M$,
\begin{eqnarray}
 \dot{t}&=&2\frac{2CE-DL}{4AC+D^{2}},\nonumber\\
 \dot{\phi}&=&2\frac{DE+2AL}{4AC+D^{2}},\label{motion}\\
 \dot{r}^{2}&=&4\frac{CE^{2}-DEL-AL^{2}}{B(4AC+D^{2})}.\nonumber
\end{eqnarray}
The radial motion of a photon can be rewritten as
\begin{eqnarray}
 \dot{r}^{2}+V_{\text{eff}}=0,\label{kk}
\end{eqnarray}
where the effective potential reads $V_{\text{eff}}=-4\frac{CE^{2}-DEL-AL^{2}}{B(4AC+D^{2})}$. Note that Eq. (\ref{kk}) is very similar to the classical equation of motion with a kinetic energy plus a potential energy. However, the dot here denotes the ordinary differentiation with respect to an affine parameter rather than the coordinate $t$.

Next, it is worth to introduce an important concept, the unstable circular null geodesic, which locates at the maximum of the effective potential $V_{\text{eff}}$. And the following conditions are required
\begin{eqnarray}
 V_{\text{eff}}|_{r=r_{c}}=0,\quad
 V_{\text{eff}}'|_{r=r_{c}}=0,\quad
 V_{\text{eff}}''|_{r=r_{c}}<0,
\end{eqnarray}
with $r_{c}$ the radius of the unstable circular null geodesics. Solving the first condition, we have the minimum angular momentum
\begin{eqnarray}
 \tilde{L}_{c}=\frac{L_{c}}{E}=\frac{-D_{c}+\sqrt{4A_{c}C_{c}+D_{c}^{2}}}{2A_{c}},
\end{eqnarray}
where all the functions take values at $r_{c}$. The second condition leads to
\begin{eqnarray}
A_{c}C'_{c}-A'_{c}C_{c}+\tilde{L}_{c}(A'_{c}D_{c}-A_{c}D'_{c})=0.
\end{eqnarray}
Considering the third condition, $r_{c}$ is found to be the largest root of this equation. For the photon coming from infinity, approaching the nearest distant $r_{0}$, and then returning back to infinity, the deflection angle can be expressed as
\begin{eqnarray}
 \alpha(r_{0})=2\int_{r_{0}}^{\infty}\frac{\dot{\phi}}{\dot{r}}dr-\pi.
\end{eqnarray}
In an asymptotically flat black hole background, the deflection angle increases with the decreasing of $r_{0}$ and gets infinity when $r_{0}$ approaches $r_{c}$. Near $r_{c}$, Bozza \cite{Bozza02} showed an approximate deflection angle,
\begin{eqnarray}
 \alpha(u)=-\bar{a}\ln\bigg(\frac{u}{u_{c}}-1\bigg)
             +\bar{b}+\mathcal{O}(u-u_{c}).
\end{eqnarray}
where the minimum impact parameter $u_{c}$, and the strong deflection limit coefficients $\bar{a}$, $\bar{b}$ are parameterized with the radius $r_{c}$. For our study, we only show the parameters $u_{c}$ and $\bar{a}$, which read
\begin{eqnarray}
 u_{c}&=&\tilde{L}_{c},\\
 \bar{a}^{2}&=&\frac{2A_{c}B_{c}}{A_{c}C_{c}''-A_{c}''C_{c}+u_{c}(A_{c}''D_{c}-A_{c}D_{c}'')}.
\end{eqnarray}
Since we have adimensionalized the black hole parameters with the mass $M$, $u_{c}$ is a dimensionless quantity. According to the lens geometry presented in Ref. \cite{Virbhadra}, there would be two infinite series of images close to the black hole. Assuming that the source, lens, and observer are highly aligned, there would be two observables. The first one is the minimum angular position $\theta_{\infty}$ of the images, and the second one is the ratio $\tilde{r}$ between the flux of the first image and the sum of the others, which is related to the relative magnitudes as $r_{m}=2.5\lg \tilde{r}$ with $\tilde{r}=e^{2\pi/\bar{a}}$. The two observables can be constructed from $u_{c}$ and $\bar{a}$
\begin{eqnarray}
 \theta_{\infty}=\frac{u_{c}}{D_{\text{OL}}}, \quad
 r_{m}=\frac{6.8219}{\bar{a}}.
\end{eqnarray}
Since $u_{c}$ and $\bar{a}$ depend on the feature of the black hole lens, this relation provides an opportunity to check the consistency between theoretical model and astronomical observations.

Now, let us turn to the QNM frequencies. As pointed out above, in the eikonal limit, the frequencies are in the form of Eq. (\ref{One}), and the parameters $\Omega_{c}$ and $\lambda$ are given by \cite{Cardoso}
\begin{eqnarray}
 \lambda=\sqrt{-\frac{V_{eff}''}{2\dot{t}^{2}}}\bigg|_{r_{c}},\quad
 \Omega_{c}=\frac{\dot{\phi}}{\dot{t}}\bigg|_{r_{c}}.
\end{eqnarray}
Supposing that the equatorial plane of the spacetime can be described by the metric (\ref{reducemetric}), and by using Eqs. (\ref{motion}) and (\ref{kk}), the parameters $\lambda$ and $\Omega_{c}$ can be expressed as \cite{wei}
\begin{eqnarray}
 \Omega_{c}&=&\frac{1}{\tilde{L}_{c}},\\
 \lambda^{2}&=&\frac{A_{c}C_{c}''-A_{c}''C_{c}+\tilde{L}_{c}(A_{c}''D_{c}-A_{c}D_{c}'')}
                    {2A_{c}B_{c}\tilde{L}_{c}^{2}}.
\end{eqnarray}
Combining with the gravitational lensing, we can express $\Omega_{c}$, and $\lambda$ in terms of $u_{c}$ and $\bar{a}$,
\begin{eqnarray}
 \Omega_{c}=\frac{c}{u_{c}R_{s}},\quad
 \lambda=\frac{c}{\bar{a}u_{c}R_{s}},\label{omega1}
\end{eqnarray}
where $R_{s}=\frac{GM}{c^{2}}$ with $G$, $c$, and $M$ the Newton gravitational constant, speed of light and mass of the black hole, respectively. Note that here we have restored the dimension. This result implies that the angular velocity at the unstable circular null geodesics is just the inverse of the minimum impact parameter. Further, using the observables from the lensing, we have
\begin{eqnarray}
 \Omega_{c}=\frac{c}{\theta_{\infty}D_{\text{OL}}},\quad
 \lambda=0.1468c\frac{r_{m}}{\theta_{\infty}D_{\text{OL}}}.\label{omega2}
\end{eqnarray}
This is our primary result. It is worth to point out that this relation is dependent of the lens geometry, i.e., the distance $D_{\text{OL}}$, while it is independent of the nature of the black hole. So with predetermined $D_{\text{OL}}$, $\Omega_{c}$ and $\lambda$ can be accurately determined by the observables $\theta_{\infty}$ and $r_{m}$ without knowing the detail of the black hole lens.

\section{Kerr-Newman black holes}
The KN black hole is the most generally vacuum solution of the Einstein-Maxwell equations, which can describe the spacetime geometry in the region surrounding a charged, rotating mass. The metric corresponding to it is in the form of (\ref{Kerr}) with the metric functions given by
\begin{eqnarray}
 &&\Delta=r^{2}-2Mr+a^{2}+Q^{2},\quad P(\theta)=-a\sin^{2}\theta,\\
 &&\rho^{2}=r^{2}+a^{2}\cos^{2}\theta,\quad R(r)=-(r^{2}+a^{2}).
\end{eqnarray}
where $a$, $Q$, and $M$ are the spin, charge, and mass of the black hole. So a KN black hole can be characterized by a pair of parameters $(a/M,\;Q/M)$. Adopting the method described above, these observables $\theta_{\infty}$ and $r_{m}$ could be obtained for a KN black hole lens. In order to observe a notable lensing, the black hole lens must be a supermassive one. In general, at the center of each galaxy, there is supposed to be a supermassive black hole with mass $M\sim 10^{5}-10^{9}M_{\odot}$, where $M_{\odot}$ is the solar mass. So the gravitational lensing by it is most easily to be observed. For example, the supermassive black hole Sgr A$^{*}$ at the center of our Milky Way is estimated to be $M=2.8\times10^{6}M_{\odot}$, and the distance between the observer and the black hole is supposed to be $D_{\text{OL}}=8.5$ \emph{kpc}. Supposing that the black hole can be described by a KN metric, the minimum angular position $\theta_{\infty}$ for this case can be calculated as
\begin{eqnarray}
 \theta_{\infty}\approx 9.8557\times10^{-6}u_{c}
 \bigg(\frac{M_{BH}/M_{\odot}}{D_{\text{OL}}/1\text{kpc}}\bigg)
       \;\; \mu\text{arcsec}.
\end{eqnarray}
\begin{figure}
\includegraphics[width=8cm]{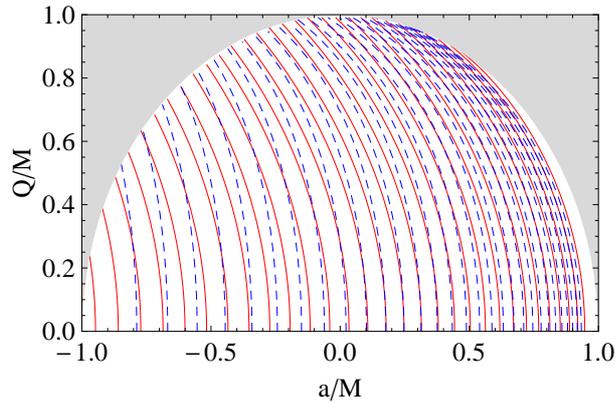}
\caption{Contour plots of the observables $\theta_{\infty}$ and $r_{m}$ in the $(a/M, Q/M)$ plane. $\theta_{\infty}$ is described by the full lines with values from $16.88\sim44.88$ $\mu$arcsec from right to left. $r_{m}$ is described by the dashed lines with values $2.2\sim 8.8$ from right to left. The light gray zone represents naked singularities.}\label{aq}
\end{figure}
For each pair of parameters $(a/M,\;Q/M)$, the observables $\theta_{\infty}$ and $r_{m}$ are uniquely determined. In the reverse way, with the values of $\theta_{\infty}$ and $r_{m}$, the black hole parameters will be determined. Therefore, in order to determine the parameters of the black hole, we show in Fig. \ref{aq} the contour curves of constant $\theta_{\infty}$ (full lines) and $r_{m}$ (dashed lines) in the plane $(a/M,\;Q/M)$. It is clear that each point in the plane is characterized by the values of both $\theta_{\infty}$ and $r_{m}$. So with the data form astronomical observations, this provides an accurate way to determine the black hole parameters.

\begin{figure}
\includegraphics[width=8cm]{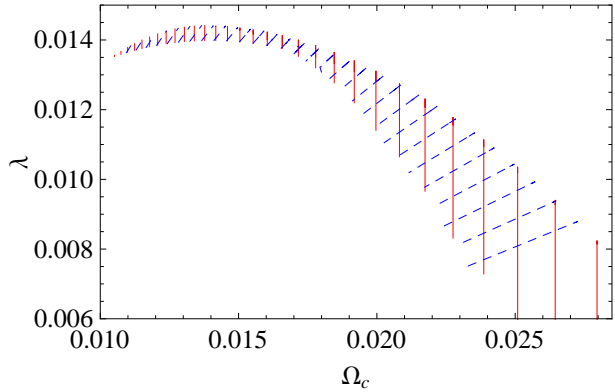}
\caption{Contour plots of the observables $\theta_{\infty}$ and $r_{m}$ in the $(\Omega_{c}, \lambda)$ plane. $\theta_{\infty}$ is described by the full lines with values from $16.88\sim44.88$ $\mu arcsec$ from right to left. $r_{m}$ is described by the dashed lines with values $2.2\sim 8.8$ from right to left.}\label{omegalambda}
\end{figure}

Next, we wish to determine the QNM frequencies from the observables $\theta_{\infty}$ and $r_{m}$. With the help of Eqs. (\ref{omega1}) and (\ref{omega2}), we plot the contour curves of the constant $\theta_{\infty}$ (full lines) and $r_{m}$ (dashed lines) in the plane $(\Omega_{c},\;\lambda)$ in Fig. \ref{omegalambda}. It is interesting to note that in the figure, the contour curves of the constant $\theta_{\infty}$ and $r_{m}$ are limited in a belt denoting a black hole lens rather than a naked singularity one. The curves of the constant $\theta_{\infty}$ are vertical lines in the figure due to the fact that $\Omega_{c}\varpropto 1/\theta_{\infty}$. Each point in this belt is characterized by the values of $\theta_{\infty}$ and $r_{m}$. Thus, with the data form astronomical observations, we can get the expected value of the angular velocity and Lyapunov exponent of the unstable circular null geodesics.

\section{Gravitational waves from astronomical observation}
The QNM frequency is a complex number, and its imaginary part will decay when the wave travels out far from the black hole. Therefore for a observer located at infinity, only the real part of the QNM frequencies is meaningful, which should be related to the GWs. Based on this conjecture, for Sgr A$^{*}$, the possible range of the GWs can be read out from Fig. \ref{omegalambda}, which approximately is
\begin{eqnarray}
 f\sim (1\sim 3)m\times 10^{-2}\; \text{Hz}.\label{bound}
\end{eqnarray}
Here $m$ is an integer. Since the relation (\ref{omega2}) is independent of the nature of the black hole lens, we see that this is a universal result for the black hole located at the center of the Milky Way.

For $m\sim 10^{2}$, we look forward to obtain the GWs in the range of several Hz. With the increase of $m$, one can get the GWs of $\sim$ kHz. If this is true, we can use the tunable resonant sensor proposed in Ref. \cite{Arvanitaki} to directly detect GWs, which bases on the optically trapped and cooled dielectric microspheres or micro-discs, and is believed to be effective in the range of 50$-$300 kHz.  On the other hand, Eq. (\ref{bound}) also gives a low bound $f_{\text{lb}}$ ($m$=1) for the GWs
\begin{eqnarray}
 f_{\text{lb}}\sim 10^{-2}\; \text{Hz}.
\end{eqnarray}
Thus, the observation of GWs with $f<10^{-2}$ Hz from the black hole at the center of the Milky Way is very frustrating.

\begin{figure}
\includegraphics[width=8cm]{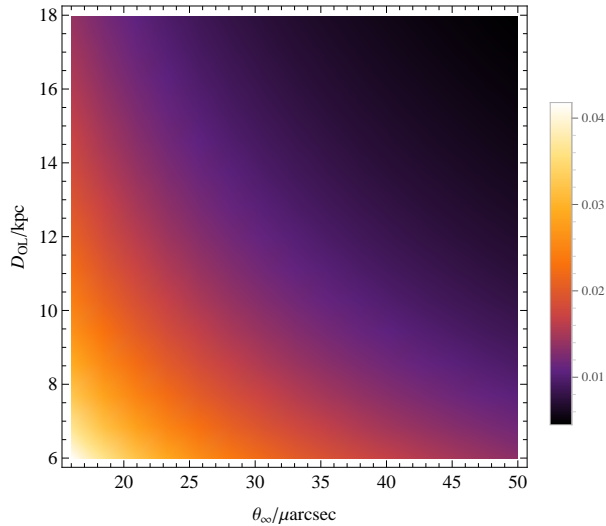}
\caption{Behavior of the low bound $f_{\text{lb}}$ for different values of $D_{\text{OL}}$ and $\theta_{\infty}$.}\label{fig3}
\end{figure}

This low bound of frequency of the GWs can also be extended to other black hole. From the relation (\ref{omega2}), we find the low bound $f_{\text{lb}}$ depends on the distance $D_{\text{OL}}$ and the observable $\theta_{\infty}$. In Fig. \ref{fig3}, we show the low bound $f_{\text{lb}}$. With $D_{\text{OL}}$ varying from 6 to 18 kpc, and $\theta_{\infty}$ from 10 to 50 $\mu arcsec$, $f_{\text{lb}}$ decreases from 0.04 to 0.005 Hz. Thus given different values of $D_{\text{OL}}$ and $\theta_{\infty}$, we can read out the value of the low bound $f_{\text{lb}}$ from this figure immediately.

\section{Conclusion}
In this letter, we have shown an elegant relation between the GWs and black hole lensing with the suppose that the black hole acts simultaneously as a GWs source and a gravitational lens. This relation is both effective for the black hole with or without the spin parameter in an asymptotically flat spacetime. It is also found to be independent of the nature of the black hole lens. Thus we could obtain the information of GWs emitted by the black hole through its lensing geometry and observable, which are more likely to be observed by modern astronomical instrument. The low bound $f_{\text{lb}}$ of the frequency of the GWs is given. Although directly detecting GWs is a challenge, the result provides a possible way to test it, or at least to locate the GWs source that most like to be observed by optical or radio telescopes with proper range of the frequency.

This work was supported by the National Natural Science Foundation of China (Grant No. 11205074 and Grant No. 11075065), the Huo Ying-Dong Education Foundation of the Chinese Ministry of Education (Grant No. 121106), and the Fundamental Research Funds for the Central Universities (Grant No. lzujbky-2013-21 and No. lzujbky-2013-18).

\end{document}